\documentclass[prd,showpacs,preprintnumbers,amsmath,amssymb,amsfonts,nofootinbib]{revtex4}
\everymath{\displaystyle}
\usepackage[dvips]{graphicx}
\usepackage{longtable}
\usepackage{subfigure}

\newcommand{\ide}{\mbox{1 \kern-.59em {\rm l}}}
\begin{document}

\preprint{CHEP-PKU/1/01-2008, IMSC/2008/01/01}

\title{Spontaneous symmetry breakdown in fuzzy spheres}

\author{C.R. Das}
\email{crdas@imsc.res.in, crdas@pku.edu.cn}
\affiliation{Center for High-Energy Physics,
Peking University, Beijing  100 871, China}
\author{S. Digal}
\email{digal@imsc.res.in}
\affiliation{Institute of Mathematical Sciences, C.I.T. Campus,
Taramani, Chennai  600 113, India}
\author{T.R. Govindarajan}
\email{trg@imsc.res.in}
\affiliation{Institute of Mathematical Sciences, C.I.T. Campus,
Taramani, Chennai  600 113, India}

\begin{abstract}
We study and analyse the questions regarding breakdown of global symmetry
on noncommutative sphere. We demonstrate this by considering a complex scalar
field on a fuzzy sphere and isolating Goldstone modes. We discuss the role of nonlocal
interactions present in these through geometrical considerations.
\end{abstract}



\maketitle

\section{Introduction}
Recently there has been interest in the non-perturbative numerical studies of 
quantum fields on fuzzy spaces. This was motivated by the fact that implementation 
of noncommutative geometry appears through novel features in such models. 
For example, the IR/UV mixing, even though absent in the finite matrix 
setting appears through an anomaly which generates it in the continuum limit. 
This suggests the possibility that fuzzy spaces may serve as novel regulators 
of field theories and bring out new features which were absent in the 
conventional lattice regularisations. This program falls well within the 
front line of research activity  of study of noncommutative geometry 
\cite{hoppe,madore,pinzul,sbal,xavier,denjoe,steinaker,bietenholz,panero,das}.

Earlier studies on simulations have focused  on scalar fields on fuzzy 
spheres ($S_F^2$) and led to the demonstration of 
new phases characterised as nonuniform
ordered phase \cite{denjoe,bietenholz,panero,das}. 
In the continuum infinite volume limit 
this nonuniform phase  appears as stripe phase in the Groenewold-Moyal 
space \cite{gubser,ambjorn}. These nonuniform phases are related to the 
existence of meta stable states and we have demonstrated this in our earlier 
analysis using what is known as pseudo heat bath method \cite{das}. We could 
also show the existence of meta stable states as well as their connections to 
many phases.

Those new ``stripe'' phases were conjectured by Gubser and Sondhi 
\cite{gubser}, where they pointed out that this translation 
non-invariant phase would exist only in dimensions 
$d\geq 3$. However the numerical simulation of Ambjorn and Catterall show 
the existence of such a phase even in $d=2$ \cite{ambjorn}. 
This seems to be contradicting Coleman-Mermin-Wagner(CMW)
theorem that continuous symmetries  cannot be broken spontaneously in $d=2$
\cite{mermin}. But
CMW theorem can be expected to be valid only in local field theories 
\cite{paolo}. On the other hand NC field theories are inherently nonlocal 
and can be expected to show new features. However Gubser and Sondhi while 
analysing NC field theories point out that the infrared problems are 
worse in these theories and hence CMW theorem cannot be
violated. But numerical simulations show the results to be other way. 
In this connection it is important to see whether long range order 
or symmetry breaking can be seen for 
global symmetries other than space time symmetries. This assumes importance,
since Gubser and Sondhi explicitly use $O(N)$ and $O(2)$ symmetries
in showing their results.

We analyse in this paper this  important question, namely what happens to 
CMW theorem and long range order in the fuzzy spaces \cite{paolo}.
We consider a complex scalar field theory with a global $U(1)$ and study 
the full implications of the noncommutative nature of the underlying geometry.
As pointed out earlier spontaneous symmetry breakdown (SSB) and long range 
order are obstructed in $2D$ theories by the infrared divergences associated 
with Goldstone modes.  While nonlocal field theories can escape these 
conditions, it is not obvious whether they do so in these circumstances.

This paper is organised as follows: In Sec.2 we describe the model, 
notations and address the questions. In Sec.3 we look at 
the aspects of simulations and demonstrate the spontaneous 
breaking of $U(1)$ symmetry thereby evading CMW theorem. 
In Sec.4, we analyse questions of Goldstone modes, the role of 
non-locality and subtle issues in isolating effects of 
continuum limit as well as infinite volume limit. In the last Sec.5,  
we present our results and conclusions, taking into account already existing 
theoretical studies.

\section{SSB on $2D$ NC space and CMW theorem}
The standard action used for studying the $U(1)$ in two dimensional ($2D$) 
Moyal spaces is given as \cite{gubser},
\begin{equation}
S = \int d^2x\left[ |\partial\Phi|^2 + r|\Phi|^2 +
\lambda_1\Phi{\bf \star}\Phi{\bf \star}\Phi^{*}{\bf 
\star}\Phi^{*} + \lambda_2\Phi {\bf \star}\Phi^*{\bf\star}\Phi{\bf\star}
\Phi^{*}\right]
\end{equation}
Finite temperature behaviour of the $\Phi$ field
with fluctuations is studied by varying the mass parameter $r$. One would 
expect the average of $\Phi$ to take the form,
\begin{equation}
\langle|\Phi|\rangle = \sqrt{-{\frac{r}{2(\lambda_1+\lambda_2)}}}\,
\end{equation}
in the mean-field theory. In commutative two dimensional space the 
$U(1)$ Goldstone mode destroys the above condensate. So, there will 
not be SSB of the $U(1)$ symmetry in these spaces. As discussed above, 
the status of $U(1)$ SSB is not clear in $2D$ NC spaces. Very little
has been done from the non-perturbative side to study this issue. 

We will present in this note that the spontaneous breaking of internal 
symmetries can be expected in Moyal space times when space-time symmetries
are broken. Conventionally the obstruction to SSB in $2D$ will come through the 
infrared divergence from the resulting massless Goldstone boson. Gubser 
and Sondhi argued that the infrared behaviour is worse in $2D$ NC space and 
hence will continue to obey CMW theorem. But the simulations point out that 
this to be not the case. We will argue how this can be understood with 
specific requirements in these theories.

To obtain SSB of $U(1)$ we should have the complex field $\Phi$
in the ground state such that 
\begin{equation}
\int\Phi\, d^2x =0\quad \rm{and}\quad \Phi \ne 0\,. 
\end{equation}

This will translate to $Tr\Phi=0$ in Moyal space. This implies that when 
$\Phi$ is expanded in Fourier modes $\tilde\Phi(k)$, the zero mode should be 
absent. That is, 
$$
\tilde{\Phi}(k=0)=0\,.
$$
Correspondingly, on the sphere $\Phi$ should not 
have a singlet component in the angular momentum basis. This naturally gives 
a cutoff $k_c$ fixed by the parameters of the effective action 
including quantum fluctuations.
This cut-off in the integral avoids infrared singularities.
The important factor responsible for this is the existence of states
which are not translationally invariant, such as the stripe phase (or
nonuniform phase). Hence $U(1)$ can be broken spontaneously only
along with translation symmetry violating nonuniform phase.
We will see how these features appear in our simulation studies.

\section{Simulations on Fuzzy $S^2$}
In this work we study this problem non-perturbatively on a fuzzy sphere.
The fuzzy spheres are described by the coordinates $X_i$ obeying the algebra
\begin{equation}
\left[X_i,X_j\right]=\frac{i\alpha \epsilon_{ij}{^{k}}X_k}{{\sqrt{j(j+1)}}},~~~
\sum_iX_i^2=R^2.
\end{equation}
Here $R$ is the radius of
the fuzzy sphere. $\alpha/\sqrt{j(j+1)}$ measures the non commutativity in
the fuzzy sphere. Here $\alpha$ is a function of $R$.
On the fuzzy sphere the above action (Eq.1) reduces to,

\begin{equation}
S(\Phi) = \frac{4\pi}{N}\,{\rm Tr}\left[
\sum_i |\left[L_i,\Phi\right]|^2
+R^2\left(r|\Phi|^2+\lambda_1\Phi\Phi\Phi^{\dag}\Phi^{\dag} + \lambda_2\Phi\Phi^{\dag}\Phi
\Phi^{\dag} \right) \right].
\end{equation}

\noindent Here the real and imaginary parts of $\Phi\in {\rm Mat}_N$ are
$N\times N$ hermitian matrices. The quartic terms represent the self 
interaction of the $\Phi$ field. These terms are crucial for the presence
of noncommutativity and nonlocality in this problem. In the following
we describe our strategy of studying the SSB of $U(1)$ and the numerical
simulations.

The simulation of the above model for the SSB study of $U(1)$ in fuzzy $S^2$
involves generating statistically relevant matrix configurations for
a particular choice of parameters. The configurations are
generated using ``pseudo-heatbath'' method, for details see \cite{das}. The
auto-correlation for larger $N$ simulations turn out to be large. To reduce
the autocorrelation problem we use over-relaxation algorithm. 
However the number of over-relaxation steps required grows with $N$ rapidly. 
This makes it difficult to get any reasonable statistics for large $N$. 

The CMW theorem and SSB for finite systems is subtle. 
For a finite system there will be tunnelling between all possible vacua, 
though the tunnelling rate is exponentially suppressed as the size of 
the system increases. Because of the tunnelling, the state of the 
system respects the symmetry. However this state of the system is unlike 
the symmetry restored state at high temperatures. In this situation
ensemble average of the magnitude of the ``magnetisation'' is usually taken 
to be the condensate (order parameter). Using this prescription 
we look for SSB of continuous symmetry for 
finite systems in $2D$.  In order to numerically prove the CMW theorem 
one has to show that the condensate vanishes in the infinite 
volume limit. The numerical study 
of finite size effects in commutative $2D$ show that the condensate vanishes 
logarithmically. One should expect that there will also be finite size 
effects for noncommutative spaces. So the finite $R,N$ effects must be 
separated out to settle the issue of $U(1)$ SSB on fuzzy sphere. There are 
various limits possible consistent with $(R,N) \to \infty$. 
We do not consider the limit $(R,N) \to \infty$
in which the resultant space becomes commutative and there 
should be no SSB of $U(1)$. When the
ratio $R^2/N$ remains finite for $(R,N)\to \infty$ we have a thermodynamic
limit which still maintains the non-commutative character.

The numerical simulations always consider a finite $N$. As discussed above
a non-zero $\langle \Phi \rangle$ will not validate or invalidate the
CMW theorem. We need to study the behaviour of $\langle\Phi\rangle$ for 
various $N$ to see SSB of $U(1)$ or otherwise. 
While we vary $N$ the ratio $R^2/N$ is kept
fixed. To simplify the numerical work we choose
$\lambda_1=\lambda_2=\lambda/2$ from the parameter space. The simulations
were then done for a fixed $\lambda,r$. 
The observables we measure are action $S$, $\Phi$ and fluctuations of various
elements of $\Phi$. Each element $\Phi_{ij}$ now is a complex number, so,

\begin{equation}
\Phi_{ij}=\eta_{ij}e^{i\theta_{ij}}.
\end{equation}

Note that because of the $U(1)$ symmetry $\Phi_{ij}$ will have a circular 
distribution in the complex plane, with $\langle \eta_{ij} \rangle$ being 
the average radius of the distribution. A non-vanishing
$\langle \eta_{ij} \rangle$ in the $N \to \infty$ limit for at least
one element of $\Phi$ will clearly signal SSB of $U(1)$. In the following
we discuss the simulation parameters, results and discussions.

\section{Results and discussions}

The parameter values we choose for our simulations are $\lambda=1.0$,
$r=-8$. We have also repeated our simulations for $r=-16$. For these choices
of parameters various values of $N$ starting from $N=16$ onwards are 
considered. The run time to get reasonable statistics grows rapidly
with $N$. At present the largest value of $N$ we have simulated is $N=100$. 
We also have a few set of data with less statistics for $N > 100$. 

The simulation results showed that the off-diagonal elements of $\Phi$
are always fluctuating around zero. The change in the behaviour of these
elements with $N$ is not so prominent. So, in this paper we show only the 
behaviour of diagonal elements of $\Phi$. In Fig.1, we show the distributions 
of $\Phi_{11},\Phi_{22}$ and $\Phi_{mm}$, where $mm$ correspond to the 
middle diagonal element or one of the middle diagonals if $N$ is even. 

\begin{figure}[hbt!]
\begin{center}
\subfigure[\;Distribution of $\Phi_{11},\Phi_{22}$ and $\Phi_{mm}$ for $N=19$
and $\rho_0 \sim 1$.]
{\label{fig1a}\includegraphics[scale=0.59]{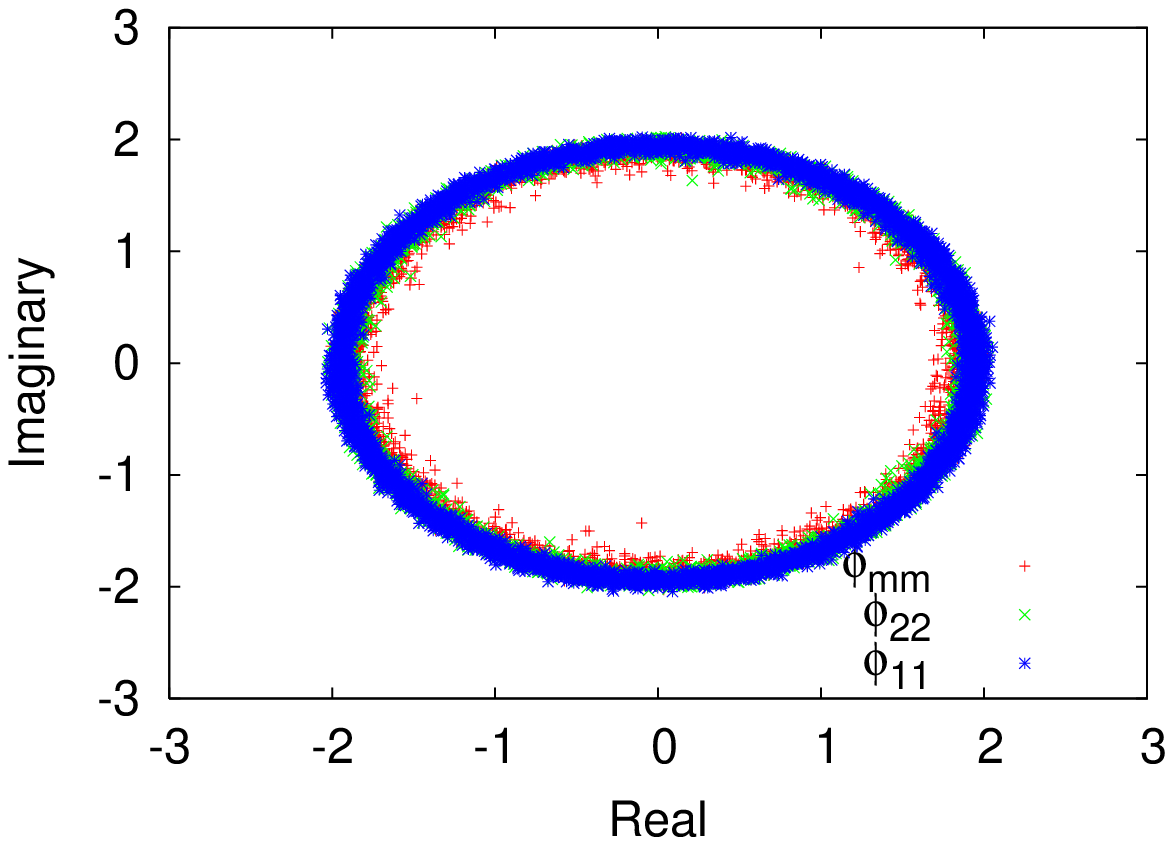}}
\subfigure[\;Distribution of $\Phi_{11},\Phi_{22}$ and $\Phi_{mm}$ for $N=64$
and $\rho_0 \sim 1$.]
{\label{fig1b}\includegraphics[scale=0.59]{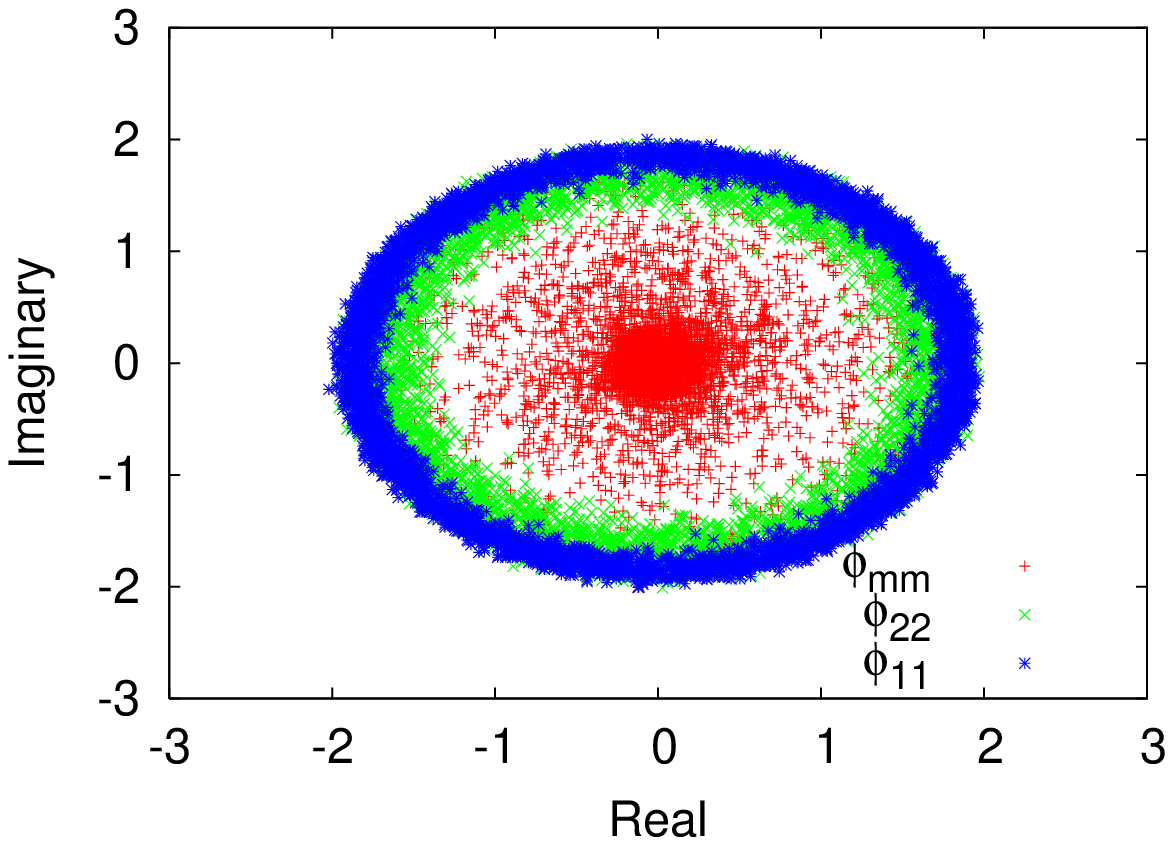}}
\caption{}
\end{center}
\end{figure}

From the figure we can clearly see that the distribution has a $U(1)$ 
symmetry. The distributions of pair of elements such as $\Phi_{ii}$ 
and $\Phi_{N-i+1,N-i+1}$ are found to be the same as expected.
In order to show finite size effects we consider $N=19$
and $N=64$ in the above figure. We see that for all $N$, the radii of
the distributions decrease as we move towards the middle elements. The
difference here is that for $N=19$ the radii are close to each other, but
for $N=64$ they are different by large amount. 
Indeed for $N=64$ the middle diagonal
distribution has a peak around zero in the complex plane. The difference
in the property of $\Phi$ between $N=19$ and $N=64$ is purely finite volume
effect. Also due to finite volume effects even the radius of distribution
of $\Phi_{11}(\Phi_{NN})$ has increased from $N=64$ to $N=19$.

One of the main difference between commutative and non-commutative $2D$ space
is the presence of degenerate/meta-stable states in the latter. In the
numerical simulations one always chooses an initial configuration. For
commutative $2D$, whatever may be the initial configuration, there is always
a unique thermalized state. The situation is completely different in
the case of fuzzy sphere. Depending on the initial configuration we get
different thermalized states. 

The detailed analysis of $\Phi$ reveals that the phase of all diagonal 
elements is perfectly correlated for smaller $N$. For example in Fig.2 we 
plot the phase of $\Phi_{11}$ vs phase of $\Phi_{NN}$ for $N=19$.

\begin{figure}[hbt!]
\begin{center}
\subfigure[\;Distribution of $\theta_{11}$ vs $\theta_{NN}$ for $N=19$ and 
$\rho_0 \sim 1$.]
{\label{fig2a}\includegraphics[scale=0.59]{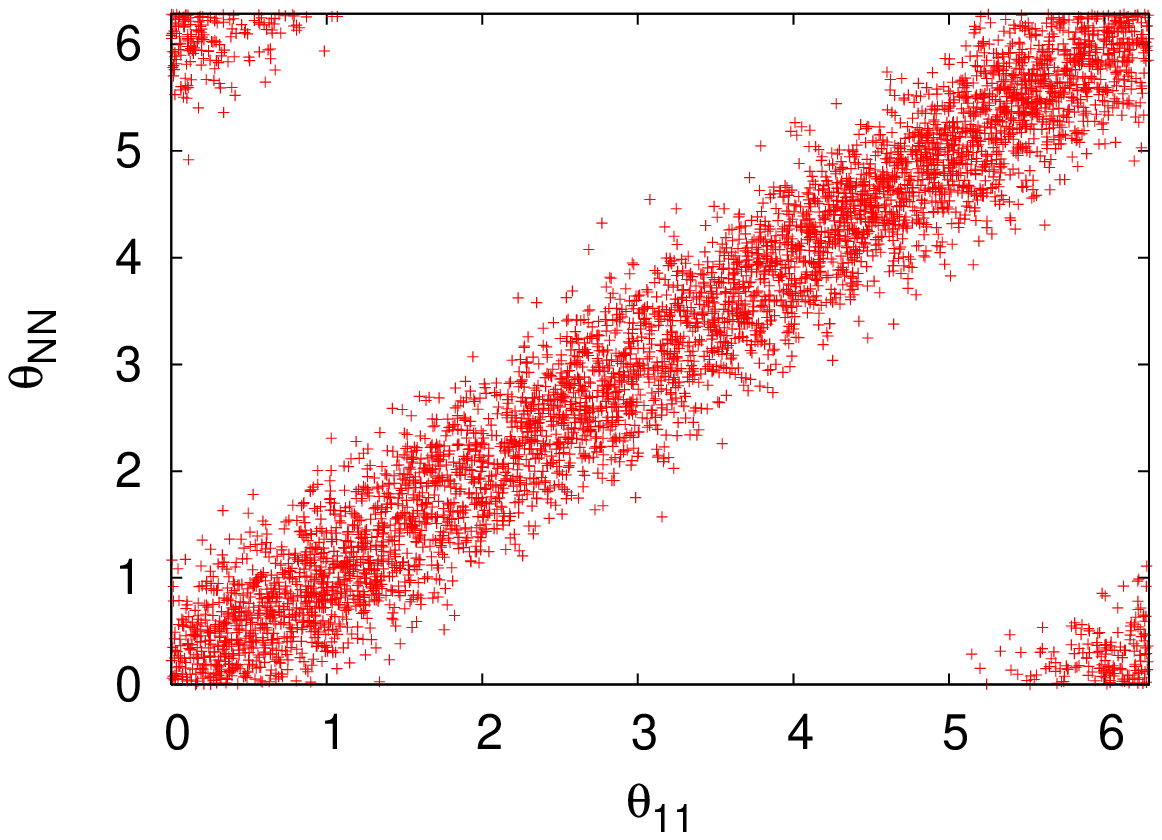}}
\subfigure[\;Distribution of $\theta_{11}$ vs $\theta_{NN}$ for $N=64$ and
$\rho_0 \sim 1$.]
{\label{fig2b}\includegraphics[scale=0.59]{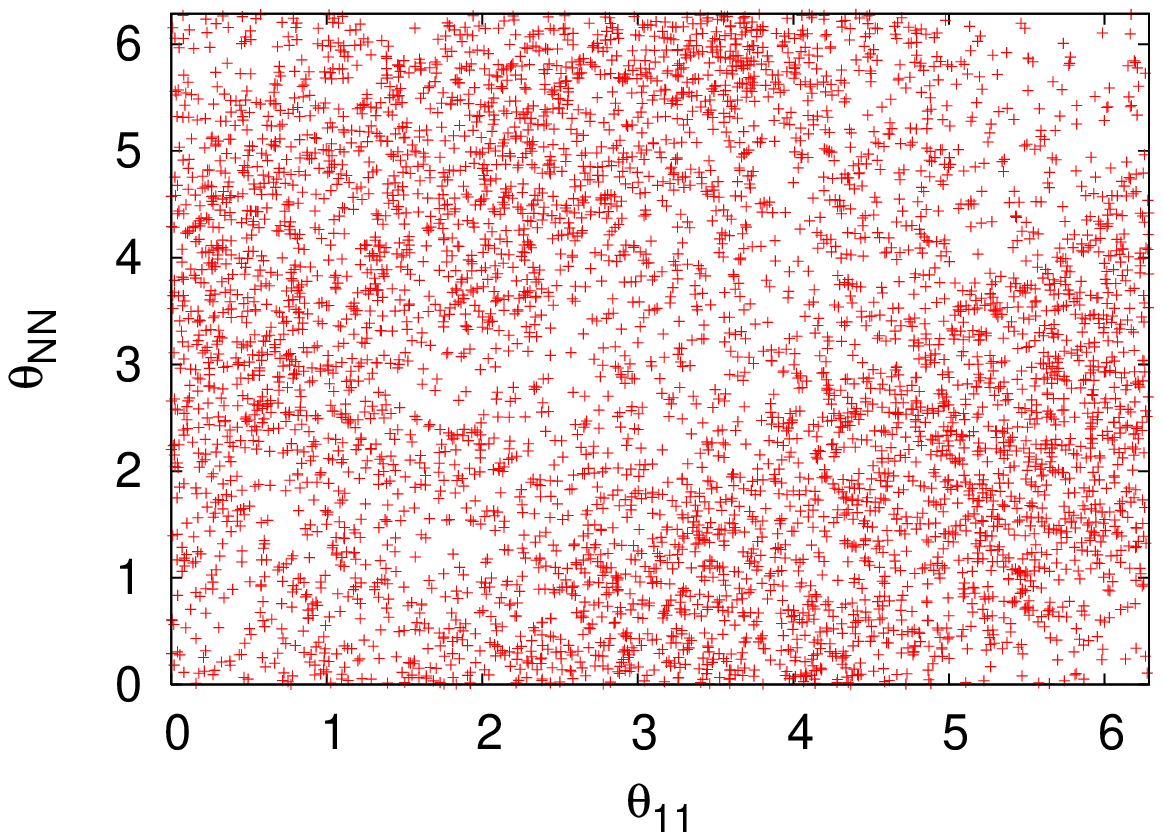}}
\caption{}
\end{center}
\end{figure}

We find that each configuration for $N=19$ the phases of all diagonal 
elements are fluctuating around same average value. Since the different 
radii for the distributions are not much different the form of $\Phi$, 
for $N=19$ is close to the form $\Phi \sim {\bf I}e^{i\theta}$. 
In
Fig.2b we plot the phase of $\Phi_{11}$ vs phase of $\Phi_{NN}$ for $N=64$.
Here, the correlation between $\theta_{11}$ and $\theta_{NN}$ is not as 
strong. These results were obtained with a initial configuration
$\Phi_0=\rho_0 {\bf 1}$, with $\rho_0 \sim 1$. With the initial condition,
$\rho_0 \sim 0$, we get same results for $N=19$ as before, but for
$N=64$ the results are dramatically different.
In Fig.3 we show the distribution of $\Phi$ and 
$\theta_{11}~\rm{vs}~\theta_{NN}$ correlation for $N=64$. 
It can be seen from Fig.3 the correlation between $\theta_{11}$ and 
$\theta_{NN}$ is stronger in this case.

\begin{figure}[hbt!]
\begin{center}
{\label{fig3b}\includegraphics[scale=0.59]{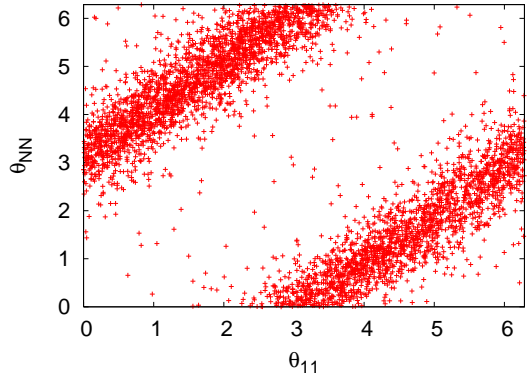}}
\caption{Distribution of $\theta_{11}$ vs $\theta_{NN}$ for $N=64$
and $\rho_0 \sim 0$.}
\end{center}
\end{figure}


The difference in results for $N=64$ shown in Fig.2b and Fig.3 
are coming from two (meta)stable states available for the system. Such 
(meta)stable states are absent for the same theory in commutative $2D$ space. 
For all different initial configurations for $N>25$, 
$\theta_{11}$ and $\theta_{NN}$ were never in phase like Fig.2a. As a
result $Tr(\Phi)$ decreases for larger $N$. For larger $N$ average of 
$Tr(\Phi)$ is found to be zero. This is expected as only these states
with $Tr(\Phi)=0$ can survive the thermodynamic limit. 

From $N=19$ to $N=64$ there is transition in the form of $\Phi$. Also the
average of $\eta_{11}$ is smaller for $N=64$. This result clearly shows
that uniform ordered phase is not a stable state in the limit $N\to \infty$.
The transition in the form of $\Phi$, when $N$ is increased takes place 
around $N=25$ for the choice of $\lambda, r$ in our simulations. This
can be seen in Fig.4a.

\begin{figure}[hbt!]
\begin{center}
\subfigure[\;$N$ dependence of $\eta_{11},\eta_{22}$ and $\eta_{mm}$.]
{\label{fig4a}\includegraphics[scale=0.59]{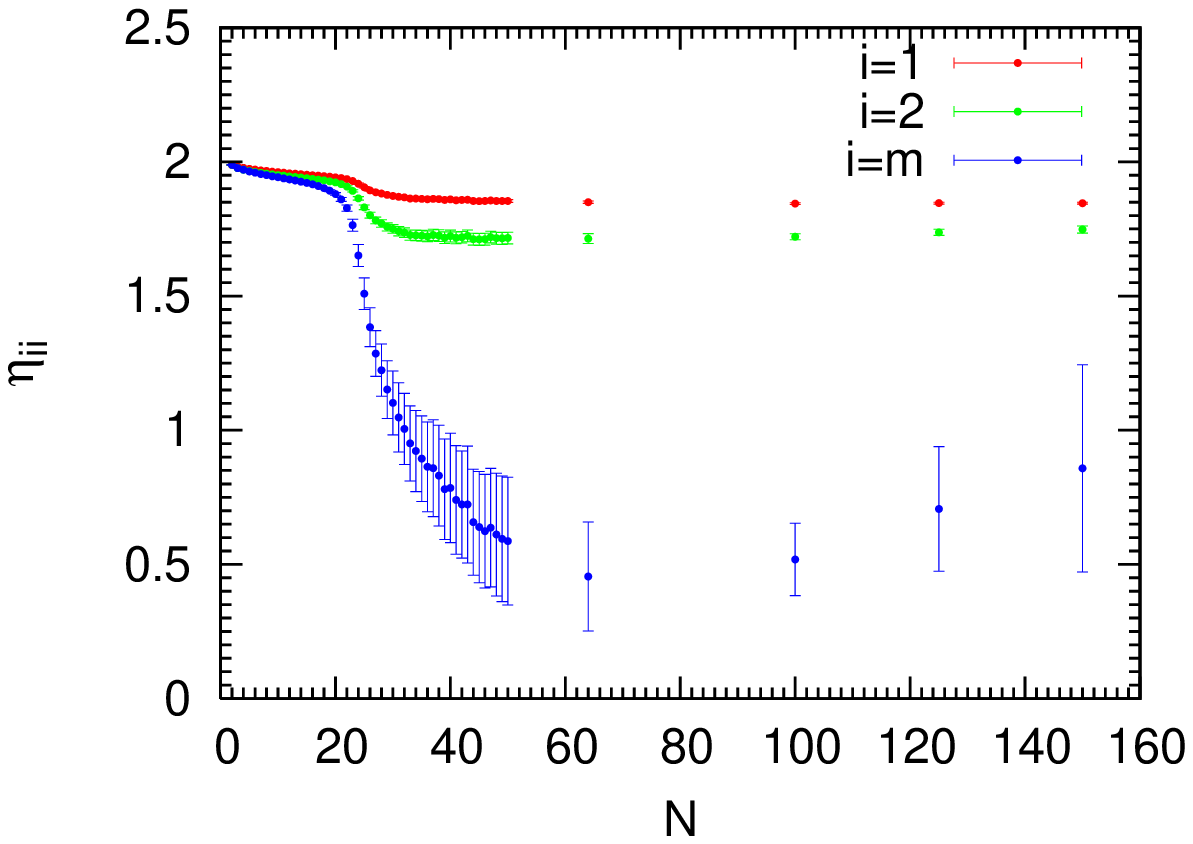}}
\subfigure[\;Histogram of $\eta_{11}$ for $N=40,64$ and $N=100$.]
{\label{fig4b}\includegraphics[scale=0.59]{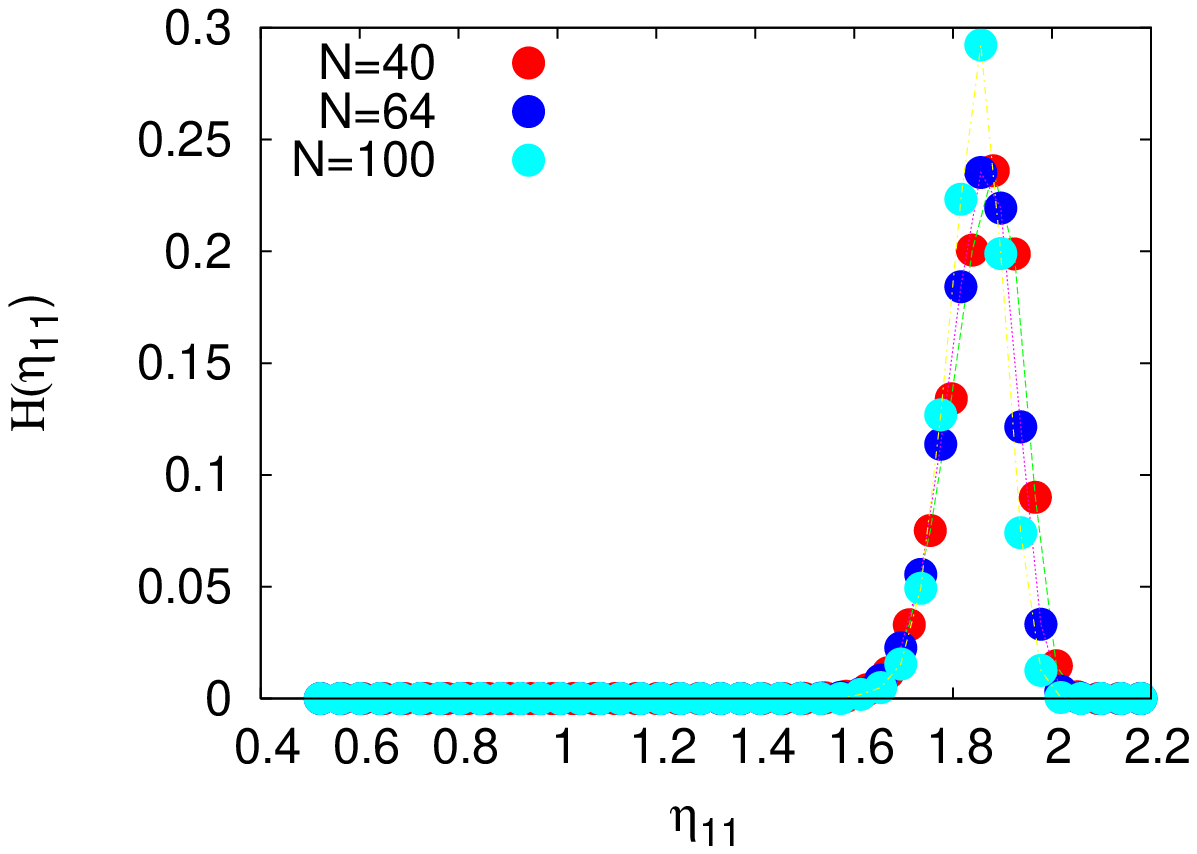}}
\caption{}
\end{center}
\end{figure}

The above results show that $\langle \eta_{11}\rangle$ seems to
saturate for $N \ge 40$. In Fig.4b, we show the histogram distribution of
$\eta_{11}$ for $N=40,64$ and $100$. The peak of the distribution hardly 
changes with $N$. On the other hand, the width for $N=100$ is clearly smaller. 
In other words, the fluctuations of $\eta_{11}$ is sharply peaked. This
suggests that the results for larger $N$ will not deviate from what we 
have already seen. The results shown in Fig.4a have been obtained with
$\rho_0 \sim 1$ for the initial configuration. We find that 
the results for $\eta_{11}$ or $\eta_{NN}$ hardly change with different
initial configurations. Even if there is a small change, their values
do not show any $N$ dependence for larger $N$. We caution, here that the 
statistics for $N > 100$ are very small and the autocorrelation is 
substantial.

The saturation of $\langle \eta_{11}\rangle$
beyond $N=40$ up to the largest simulated $N$ suggests that $\Phi$ does not
vanish in the non-commutative thermodynamic limit. The form of $\Phi$
we have obtained from the simulations also show that the ground state
is not a uniform ordered phase, but rather a non-uniform ordered
phase. This shows that all the generators, $L_x,L_y$ and $L_z$ are
spontaneously broken in the ground state. This implies that there will be
three Goldstone modes. 

\section{Conclusion} 

Having presented various simulations on fuzzy
sphere with the intention to understand the issue of CMW theorem 
we will now summarise the conclusions. While considering QFT's in 
Moyal space time Gubser and Sondhi \cite{gubser} concluded the possibility 
of translation symmetry violating stripe phase in dimensions $d\ge 3$. 
Ambjorn and Catterall \cite{ambjorn} have demonstrated this phase in
$d=2$ through Monte Carlo simulations. This was followed by studies on
$S_F^2$ and $S_F^2 \times R$ \cite{panero,bholz,medina}.
Gubser and Sondhi also conjectured that internal symmetries like 
$O(N)$ cannot be broken, as expected from CMW theorem. But CMW theorem
explicitly uses locality of interactions and NC field theories 
are inherently nonlocal. We have argued that the Fourier 
transform of the complex field should be vanishing at zero momentum,  
as the important criterion for global symmetry breaking phases. This 
translates to vanishing of the singlet component of the field in the 
angular momentum basis in fuzzy spheres. In our simulations we find 
evidence for the breaking of global symmetries in $2D$ NC field theories. The 
existence of nonuniform stable states contribute at higher temperature, 
thereby breaking the translation and internal symmetries. Translation 
symmetry breaking stripe phase is required for obtaining the internal
symmetry violation and long range order and these two occur together.
\vskip-3.0cm
\section{Acknowledgment}
We thank A. P. Balachandran, W. Bietenholz and M. Panero for useful 
discussions and comments on the draft of this paper. We also thank G. Menon 
and P. Ray for clarifying issues related to the CMW theorem. 
We also thank A. Bigarini for sending copy of his thesis where SSB and CMW 
theorem in Moyal space are discussed. 
All the calculations have been carried out using
the computer facilities at IMSc, Chennai.

\end{document}